\documentclass[12pt, reqno]{amsart}
\usepackage{amsmath}
\usepackage{amsthm}
\usepackage{amssymb}
\usepackage{amsmath}
\usepackage{amsthm}
\usepackage{amssymb}
\usepackage{fancyhdr}
\usepackage{enumerate}
\usepackage{amscd}
\usepackage[all]{xy}
\usepackage{graphicx}

\theoremstyle{remark}

\newtheorem*{exang}{Examples of divergent angular momentum}
\newtheorem*{exctm}{Examples of divergent center of mass}

\newtheorem*{pf}{Proof}
\newtheorem*{PF}{Proof of Theorem \ref{thm:main}}
\newtheorem*{pfc}{Proof of the claim}

\theoremstyle{plain}
\newtheorem{THM}{Theorem}

\newtheorem{lemma}{Lemma}[section]
\newtheorem{prop}[lemma]{Proposition}

\newtheorem{defi}[lemma]{Definition}
\newtheorem*{rmk}{Remark}
\newtheorem{cor}[lemma]{Corollary}
\numberwithin{equation}{section}

\newcommand{\og}{\overline{g}} 
\newcommand{\op}{\overline{\pi}}
\newcommand{\wg}{\widetilde{g}}

\renewcommand{\P}{{\bf P}}
\newcommand{\C}{{\bf C}}
\newcommand{\J}{{\bf J}}

\begin{document}
\title{Solutions of Special Asymptotics to the Einstein Constraint Equations}
\author{Lan--Hsuan Huang}
\thanks{The author was partially supported by the NSF through DMS-1005560.}

\address{Department of Mathematics\\
		 Columbia University\\
		 New York, NY~10027}
\email{lhhuang@math.columbia.edu}

\begin{abstract}
We construct solutions with prescribed asymptotics to the Einstein constraint equations using a cut-off technique. Moreover, we give various examples of vacuum asymptotically flat manifolds whose center of mass and angular momentum are ill-defined.
\end{abstract}
\maketitle
\pagestyle{myheadings}
\markright{Solutions of Special Asymptotics to the Einstein Constraint Equations}

\section{Introduction}
Let $M$ be a three-dimensional manifold. Let $g$ be a Riemannian metric and $K$ be a symmetric $(0,2)$--tensor on $M$. The Einstein constraint equations are
\begin{align*}
	R_g - | K|_g^2 + (\mbox{tr}_g K )^2 &= 2 \mu,\\
	 \mbox{div}_g (K - (\mbox{tr}_g K )g ) &= J,
\end{align*}
where $\mu$ and $J$ are energy density and momentum density respectively. The triple $(M, g, K)$ is called an initial data set if it satisfies the above constraint equations. It is called a \emph{vacuum} initial data set if additionally $\mu=0$ and $J=0$. In general relativity, the constraint equations are derived from the Gauss and Codazzi equations for the hypersurface $M$ in spacetime satisfying the Einstein equation, and $g$ and $K$ are respectively the induced metric and the induced second fundamental form of $M$ (see, for example \cite{W}).

An initial data set $(M,g,K)$ is called \emph{asymptotically flat} at the decay rate $q$ if, outside a compact set, $M$ is diffeomorphic to $\mathbb{R}^3 \setminus B_1$ and if there exists an asymptotically flat chart $\{x\}$ so that, for some $q>1/2$, 
\begin{align*}
		g_{ij}(x) = \delta_{ij} + O_2(r^{-q}), \quad K_{ij}(x) = O_1(r^{-1-q})
\end{align*}
and 
\begin{align*}
		\mu(x) = O(r^{-2-2q}), \quad J (x) = O(r^{-2-2q}), 
\end{align*}
where $r = \sqrt{x_1^2+ x_2^2 + x_3^2}$ and the subscript in the big $O$ notation indicates the corresponding decay rate on successive derivatives, e.g. $f = O_1(r^{-q})$ means that there is a constant $C$ uniformly in $r$ so that $|f| \le C r^{-q}$ and $|\partial f| \le C r^{-1-q}$. 

We introduce the momentum tensor $\pi = K - (\mbox{tr}_g K )g$, and define the constraint map
\[
	\Phi(g, \pi) = \left(R_g - | \pi|_g^2 +\frac{1}{2} (\mbox{tr}_g \pi )^2 , \mbox{div}_g \pi \right).
\]
Then the Einstein constraint equations take the form $\Phi(g, \pi) = (2\mu, J)$.

In the asymptotically flat chart, the following physical quantities are defined as limits of surface integrals over Euclidean spheres with the standard surface measure on $\{ r = \rho \}$:
\begin{align}
		E &= \frac{1}{16\pi} \lim_{\rho \rightarrow \infty} \int_{r= \rho} \sum_{i,j}(g_{ij,i} - g_{ii,j}) \frac{x_j}{r} \, dS,\label{def:mass}\\
		\P_i & =  \frac{1}{8\pi} \lim_{\rho \rightarrow\infty } \int_{r= \rho} \sum_{j} \pi_{ij} \frac{x_j}{r} \, dS, \label{def:lm}\\
		\C_l &= \frac{1}{16 \pi E} \lim_{\rho \rightarrow \infty} \int_{r=\rho}   \left[x_l \sum_{i,j}   ( g_{ij,i} - g_{ii,j} ) \frac{x_j}{r} - \sum_{i}  \left( g_{il}\frac{x_i}{r} - g_{ii} \frac{x_l}{r}\right)\right] \,dS,\label{def:ctm}\\
		\J_i & = \frac{1}{8\pi} \lim_{\rho \rightarrow \infty} \int_{r =\rho} \sum_{j,k} \pi_{jk}Y_{(i)}^j \frac{x_k}{r} \, dS,\label{def:am}
\end{align} 
where $Y_{(i)}$ are the rotation vector fields, e.g. $Y_{(1)} = x_3 \partial_2 - x_2 \partial_3$. These integrals correspond to the energy $E$, linear momentum  ${\P}$, center of mass ${\C}$, and angular momentum ${\J}$. 

It is well-known that the energy and linear momentum of an asymptotically flat manifold are well-defined \cite{Bartnik, C}. However, center of mass and angular momentum are the terms in the expansion of the solution $(g,\pi)$ which are of lower order than those which determine the energy and linear momentum, and they may not be well-defined unless some extra condition (for example, the RT condition below) is imposed. 

\begin{defi}
$( M, g, K )$ is asymptotically flat satisfying the Regge-Teitelboim condition (the RT condition) if $(M, g, K)$ is asymptotically flat, and $g, K$ satisfy these asymptotically even/odd conditions:
\begin{align*}
	g^{odd}_{ij} (x) = O_2(r^{-1-q}), \quad K^{even}_{ij}(x) = O_1(r^{-2-q})
\end{align*}
and 
\[
	\mu^{odd}(x) = O(r^{-3-2q}), \quad J^{odd}(x) = O(r^{-3-2q}),
\]
where $f^{odd}(x) = f(x) - f(-x)$ and $f^{even}(x) = f(x)  + f(-x)$ are respectively the even and odd parts of $f$ with respect to the fixed asymptotically flat chart $\{x\}$.
\end{defi}

Assuming $E\neq 0$, the center of mass and angular momentum are well-defined for asymptotically flat manifolds with the RT condition \cite{BO, H, RT}. Moreover, all known exact solutions to the constraint equations satisfy the RT condition. In particular, the families of Schwarzschild and Kerr solutions which are exact solutions to the vacuum constraint equations satisfy the RT condition. It was not clear whether the vacuum asymptotically flat manifolds \emph{without} the RT condition do exist, because one may tend to think the asymptotics of the solutions to the vacuum constraint equations are rigid. We show that the asymptotics are not rigid. Indeed, we can construct solutions with prescribed asymptotics, in particular without the RT condition.  

Using a cut-off technique of Corvino and Schoen \cite{CS06}, we have the following theorem. Assume that $\sigma, \tau$ are symmetric $(0,2)$--tensors defined outside a compact set in $\mathbb{R}^3$. Assume that $\sigma_{ij}, \tau_{ij}$ are components of $\sigma, \tau$ with respect to the standard Euclidean coordinate chart $\{x\}$.  
\begin{THM} \label{thm:main}

Assume that $\sigma$ and $\tau$ satisfy the linearized constraint equations outside a compact set in $\mathbb{R}^3$, i.e.
\begin{align}
		\sum_{i,j} (\sigma_{ij,ij} - \sigma_{ii,jj} ) &= 0, \label{eq:sigma}\\
		\sum_i \tau_{ij,i} &= 0, \quad \mbox{for } j = 1,2,3. \label{eq:tau}
\end{align}
Furthermore, assume that $\sigma_{ij} (x) = O_2(r^{-q}), \tau_{ij}(x) = O_1(r^{-1-q})$ for $q \in (1/2,1)$. Then given any asymptotically flat initial data set $(g,\pi)$ at the decay rate greater than or equal to $q$ and $\Phi(g, \pi) = (2\mu, J)$, there exists an asymptotically flat initial data set $(\overline{g},\overline{\pi})$ with $\Phi(\og, \op) = (2\mu, J)$ so that, for some constants $A$ and $B_i, i=1,2,3$, 
\begin{align}
		\overline{g}_{ij} &= \left(1 + \frac{A} {r} \right)\delta_{ij} + \sigma_{ij} +  O_2(r^{-1-q}), \label{eq:gasymptotics}\\
		\overline{\pi}_{ij} &= \tau_{ij} + \frac{1}{r^3} \left[- B_i x_j - B_j x_i + \sum_l (B_l x_l) \delta_{ij} \right] + O_1(r^{-2-q}), \label{eq:piasymptotics}
\end{align}	
and $(g,\pi)$ and $(\overline{g}, \overline{\pi})$ are close (in the sense of weighted Sobolev spaces).
\end{THM}

In order to construct solutions of special asymptotics, one has to find explicit $\sigma$ and $\tau$ satisfying \eqref{eq:sigma} and \eqref{eq:tau}. In section \ref{sec:special}, we give examples of $\sigma$ and $\tau$. Therefore, we can construct families of asymptotically flat manifolds without the RT condition and show that their enter of mass and angular momentum \eqref{def:ctm} and \eqref{def:am} are ill-defined  (Corollaries \ref{co:ang}, \ref{co:ctm}, and \ref{co:ang2}). It is desirable to weaken the RT condition in order to define center of mass and angular momentum. The examples in section \ref{sec:special} may help us to understand these physical quantities. Another application of the above theorem when $\sigma$ and $\tau$ are chosen compactly supported in the annular shell is in the forthcoming paper of Rick Schoen, Mu--Tao Wang, and the author \cite{HSW}. \\

\noindent {\bf Acknowledgments.}
I would like to thank Rick Schoen and Mu--Tao Wang for helpful discussions. 

\section{Constructing Solutions of Prescribed Asymptotics}
To prove Theorem \ref{thm:main}, we introduce the weighted Sobolev spaces. 
\begin{defi}[Weighted Sobolev Spaces]
For an integer $k \ge 0$, a real number $p \ge 0$, and a real number $q$, we say $ f \in W^{k,p}_{-q}$ if 
\begin{eqnarray*}
		\| f\|_{W^{k,p}_{-q}} \equiv \left( \int_M \sum_{|\alpha| \le k} \left( \big| D^{\alpha} f \big| \xi^{ | \alpha| + q }\right)^p \xi^{-3} \, d \textup{vol}_g  \right)^{\frac{1}{p}} < \infty,
\end{eqnarray*} 
where $\alpha$ is a multi-index and $\xi$ is a continuous function with $\xi = |x|$ when the asymptotically flat chart is defined. When $p = \infty$, 
\begin{eqnarray*} 
		\| f\|_{W^{k,\infty}_{-q}} \equiv \sum_{|\alpha | \le k} ess \sup_{M } | D^{\alpha} f | \xi^{ |\alpha| + q}.
\end{eqnarray*} 
\end{defi}	
Then we can weaken the definition of asymptotically flat manifolds and define $(g, \pi)$ to be asymptotically flat at the decay rate $q$ if
\[
	(g-\delta, \pi) \in  W^{2,p}_{-q} \times W^{1,p}_{-1-q}
\]
and $\Phi(g, \pi) = (2 \mu, J)$ with 
\[
	(\mu,J) \in W^{0,p}_{-2-2q} \times W^{0,p}_{-2-2q}.
\]
In the proof, we assume that $p>3/2$ and $q\in(1/2,1)$. The following proof is similar to the argument by Corvino and Schoen \cite[Theorem 1]{CS06}. The difference is that they consider the case where $\sigma=0$ and $\tau=0$, and they work on \emph{vacuum} initial data sets, i.e. $\mu=0$ and $J=0$. In our case, we allow more general $\sigma$ and $\tau$, as long as they satisfy the linearized constraint equations. Their argument can apply to our setting, except that linearized equations are slightly different due to the presence of $\sigma, \tau, \mu$ and $J$.

\begin{PF}
Let $\{\phi_k\}$ be a sequence of smooth cut-off functions:
\[
		\phi_k (x) = \left\{\begin{array}{ll} 1 & \mbox{ in } B_k, \\
				\mbox{between $0$ and $1$} & \mbox{ in }  B_{2k}\setminus B_k,\\
 				0 & \mbox{ outside } B_{2k} .
		\end{array} \right. 
\]
Also $\{ \phi_k \}$  is chosen so that $\phi_k(x) = \phi_k(r)$ and  $ | \partial \phi_k | \le C/k $ and $ | \partial^2 \phi_k | \le C/k^2$ for some constant $C$ independent of $k$. 

Let $(g, \pi)$ be a given asymptotically flat manifold at the decay rate $q$. Using the cut-off technique, we consider 
\begin{align}
	\hat{g}_k &= \phi_k g + (1 - \phi_k) ( \delta+\sigma), \label{eq:ghat}\\
	\hat{\pi}_k &= \phi_k \pi+ (1 - \phi_k )\tau. \label{eq:pihat}
\end{align}
For the moment, we work on a fixed $k$ and suppress the subscript $k$ when it is clear from context.

By \eqref{eq:sigma}, \eqref{eq:tau}, and the properties of $\phi_k$, 
\[
		\Phi(\hat{g}, \hat{\pi}) = (2\hat{\mu}, \hat{J}),
\]
where $(\hat{\mu}, \hat{J}) = (\mu, J)$ in $B_k$ and $(\hat{\mu}, \hat{J}) = (O(r^{-2-2q}), O(r^{-2-2q}))$ outside $B_{2k}$ because of \eqref{eq:sigma} and \eqref{eq:tau}.

In order to fully solve the constraint equations with the given $\mu$ and $J$, we consider a function $u$ and a vector field $X$ so that 
\begin{align}
		\overline{g} &= u^4 \hat{g}, \label{eq:goverline}\\
		\overline{\pi} &= u^2 (\hat{\pi} + \mathcal{L}_{\hat{g}} X), \label{eq:pioverline} 
\end{align}
where $\mathcal{L}_{g}X$ denotes the modified Lie derivative $\mathcal{L}_{g}X= L_{g} X - (\mbox{div}_{g} X) g$ for a Riemannian metric $g$. 

\noindent {\bf Claim}: There exists $(u_k, X_k) $ with $(u_k-1, X_k) \in W^{2,p}_{-q} \times W^{2,p}_{-q}$, and $(h_k, w_k) \in W_{-q}^{2,p}\times W^{1,p}_{-1-q}$ with the compact supports such that 
\begin{align}
	\Phi(\overline{g}_k + h_k, \overline{\pi}_k + w_k) = (2\mu, J) \label{eq:constraint}
\end{align}
for $k$ large.
\begin{pfc}
To see the argument works for general $\mu$ and $J$, we only highlight the different part of the argument from \cite{CS06}.

Let $T_k : (1+ W^{2,p}_{-q}) \times W^{2,p}_{-q} \rightarrow W^{0,p}_{-2-q} \times W^{0,p}_{-2-q}$ be defined by 
\[
	T_k (u,X) = \Phi (u^4 \hat{g}_k, u^2 (\hat{\pi}_k + \mathcal{L}_{\hat{g}_k} X))
\]
be a sequence of operators. The linearization $D(T_k)_{(1,0)} $ is a Fredholm operator with index $0$ for $k$ large. Using the surjectivity of $D\Phi_{(\hat{g}_k, \hat{\pi}_k)}$, one can define 
\[
	\overline{T}_k\left((u,X),(h,w)\right) = \Phi \left(u^4 \hat{g}_k + h, u^2 (\hat{\pi}_k + \mathcal{L}_{\hat{g}_k} X) + w \right)
\]
where $(h,w)$ is chosen so that $D\Phi_{(\hat{g}_k, \hat{\pi}_k)} (h,w)$ is in the cokernel of $D(T_k)_{(1,0)}$. Therefore, $D(\overline{T}_k)_{((1,0),(0,0))}$ is an isomorphism for each large $k$ by construction. Then by applying the inverse function theorem,  $\overline{T}_k$ is a diffeomorphism from a neighborhood of $((1,0),(0,0))$ to a \emph{fixed} (independent of $k$) neighborhood of $\overline{T}_k((1,0),(0,0))$ when $k$ large. Then the image contains $(2 \mu,J)$ when $k$ large, and hence, we can find the sequence of initial data sets so that \eqref{eq:constraint} holds.
\qed
\end{pfc}
It remains to check that $(\overline{g}, \overline{\pi})$ has the desired asymptotics \eqref{eq:gasymptotics} and \eqref{eq:piasymptotics}. Fix $k$. Outside a large compact set, if we denote $\wg = \delta + \sigma$ and $\mathcal{L} = \mathcal{L}_{\wg}$, we have, by \eqref{eq:ghat}, \eqref{eq:pihat}, \eqref{eq:goverline}, and \eqref{eq:pioverline},
\begin{align*}
	\overline{g} &= u^4 \wg,\\
	\overline{\pi}& = u^2 (\tau + \mathcal{L}X).
\end{align*}
Because $(\overline{g}, \overline{\pi})$ satisfies the constraint equations, we can derive the differential equation for $u$:
\begin{align*}
		&- 8  \Delta_{\wg} u + \left[ R_{\wg}  -  | \tau |^2_{\wg} + \frac{1}{2}  \left( \mbox{tr}_{\wg} \tau \right)^2\right] u \\
		&+ u \left[  \mbox{tr}_{\wg}(\tau) \mbox{tr}_{\wg} (\mathcal{L} X )+ \frac{1}{2} ( \mbox{tr}_{\wg}(\mathcal{L} X) )^2 \right]  \\
		& - u \left[\sum_{i,j,k,l} \wg^{ij} \wg^{kl}  2 \tau_{ik} (\mathcal{L} X )_{jl} + \left| \mathcal{L} X  \right|_{\wg}^2 \right] = 2 u^5 \mu,
\end{align*}
and the differential equations for $X_i$:
\begin{align*}
	 	\mbox{div}_{\og} \op & = u^{-4} \wg^{jk} \left[ u^2 (\tau + \mathcal{L}X)_{ij;k} + 2 u u_{,k} (\tau + \mathcal{L} X)_{ij}\right] \\
		& = u^{-2} \mbox{div}_{\wg} \tau + u^{-2} \mbox{div}_{\wg} (\mathcal{L} X) + 2 u^{-3} u_{,k} \wg^{jk} (\tau + \mathcal{L} X)_{ij}\\
		& = J.
\end{align*}
Then, for $r$ large,  $u$ and $\{ X_i \}_{i=1}^3$ satisfy the differential equations of the Euclidean Laplacian $\Delta$:
\begin{align*}
		\Delta u= O (r^{-2-2q}), \quad \mbox{and }\quad \Delta X_i = O(r^{-2-2q} ).
\end{align*}
Therefore, $u$ and $X_i$ are harmonic up to lower terms; that is,
\begin{align*}
	u &= 1 + \frac{A}{4r} + O(r^{-2q}),\\
	X_i &= \frac{B_i}{r} + O(r^{-2q}),
\end{align*}
for some constants $A$ and $\{B_i\}_{i=1}^3$. Then \eqref{eq:gasymptotics} and \eqref{eq:piasymptotics} follow.
\qed
\end{PF}


\section{Solutions of Special Asymptotics}\label{sec:special}
In order to construct solutions of explicit asymptotics, the key is to solve $\sigma$ and $\tau$ in \eqref{eq:sigma} and \eqref{eq:tau}. In this section, we show some special examples of $\sigma$ and $\tau$ and use Theorem \ref{thm:main} to construct asymptotically flat initial data sets which violate the RT condition. In subsection \ref{subsec:1}, we discuss a family of solutions at decay rate exactly equal to one. In  subsection \ref{subsec:2}, we discuss another family of solutions whose decay rate is less than $1$. Moreover, we show that the center of mass and angular momentum of these examples are ill-defined.  


\subsection{Solutions at the decay rate equal to $1$}\label{subsec:1}
Let $\sigma$ and $\tau$ be
\begin{align}
	 \sigma_{ij} &= \frac{\alpha}{r} \left( \frac{x_i x_j } { r^2 } - \frac{1}{2} \delta_{ij}  \right),\label{eq:exsigma}\\
	  \tau_{ij} &= \frac{\beta}{r^2}  \frac{x_ix_j}{r^2},\label{eq:extau}
\end{align}
where $\alpha$ and $\beta$ are arbitrary $C^2$ functions defined over the unit sphere $S^2$. Because $\alpha$ and $\beta$ are independent of $r$, by direct computations, we have the following lemma:
\begin{lemma}
For any $\alpha, \beta \in C^{2}(S^2)$, $\sigma$ and $\tau$ satisfy the linearized constraint equations \eqref{eq:sigma} and \eqref{eq:tau}.
\end{lemma}

\begin{prop} \label{prop:ex}
For any $\alpha, \beta \in C^{2}(S^2)$, there exists a vacuum initial data set $(\overline{g}, \overline{\pi})$ with the following asymptotics:
\begin{align}
		\og_{ij} &= \left( 1 + \frac{A}{r}\right)\delta_{ij} +  \frac{\alpha}{r} \left( \frac{x_ix_j}{r^2} - \frac{1}{2} \delta_{ij} \right)  + O_2(r^{-1-q}),\label{eq:og}\\
		\op_{ij} &= \frac{\beta}{r^2} \frac{x_i x_j}{ r^2} + \frac{1}{r^3} \left[ -B_i x_j - B_j x_i + \sum_l (B_l x_l)  \delta_{ij} \right] + O_1(r^{-2-q}), \label{eq:op}
\end{align}
\end{prop}
\begin{pf}
The proposition follows by choosing $(g,\pi) = (\delta, 0)$ and $(\sigma,\tau)$ as \eqref{eq:exsigma} and \eqref{eq:extau} in Theorem \ref{thm:main}.  
\qed
\end{pf}
\begin{rmk}
Asymptotically flat manifolds of the above asymptotics have been discovered by Beig and \'{O} Murchadha \cite{BO}. They showed that $(\overline{g}, \overline{\pi})$ satisfies the vacuum constraint equations up to leading order terms by direct computations. Here, we provide a more rigorous treatment and prove that $(\overline{g}, \overline{\pi})$ indeed satisfies the vacuum constraint equations. 
\end{rmk}

\begin{exang}
We can construct the asymptotically flat manifolds whose angular momentum with respect to a rotation vector field $Y$ diverges.

Let $(\og, \op)$ be an asymptotically flat manifold of the asymptotics \eqref{eq:og} and \eqref{eq:op}. Fix $\rho_0$ and let $A_{\rho}= \{ x \in \mathbb{R}^3 : \rho_0 \le |x| \le \rho \}$.
\begin{lemma}\label{lemma:angularmomentum}
For any $\alpha, \beta \in C^2 (S^2)$, 
\begin{align} 
	&\int_{\partial A_{\rho}}\sum_{i,j} \op_{ij}Y^i\frac{x_j}{r} \, dS = \frac{1}{4} \int_{A_{\rho} } \sum_p\frac{\alpha_{,p}\beta}{r^3}Y^p \, dx - \int_{A_{\rho}} \frac{\alpha}{r^4} \sum_{i,j} Y^i_{,j} B_j x_i \, dx\notag\\
	&+ \int_{A_{\rho}} \left(- \frac{3 }{4 r^3} \sum_{i,p}( B_ix_i \alpha_{,p} Y^p ) - \frac{ \alpha}{r^4} \sum_p  B_pY^p\right)\, dx + O(1), \label{eq:ang}
\end{align}
where and in the following $O(1)$ denotes the term bounded uniformly in $\rho$.
\end{lemma}
\begin{rmk}
If $\alpha$ is even,  i.e. $\alpha(x) = \alpha (-x)$, then only the first integral on the right hand side contributes, and other integrals vanish. In particular, if $\alpha$ is a constant, then the angular momentum is finite no matter what choices of $\beta$ are made. 
\end{rmk}
\begin{pf}
We compute the angular momentum \eqref{eq:ang} over the annulus. Notice that because $\og = \delta + O(r^{-1})$,
\[
	\int_{\partial A_{\rho}}\sum_{i,j} \op_{ij}Y^i\frac{x^j}{r} \, dS = \int_{\partial A_{\rho}}\sum_{i,j} \op_{ij}Y^i\nu^j \, dS_{\og} +O(1).  
\]
That is, without the RT condition, the integral computed with respect to the flat metric and the one with respect to the physical metric $\bar{g}$ differ by a finite constant. Then by the divergence theorem, 
\begin{align}
			&\int_{\partial A_{\rho}}\sum_{i,j} \op_{ij}Y^i\nu^j \, dS_{\og} = \int_{A_{\rho} } \sum_{i,j,k} \og^{jk} \op_{ki} Y^i_{;j} \, d\textup{vol}_{\og} \notag\\
			&= \int_{A_{\rho} } \sum_{i,j,k} \og^{jk} \op_{ki} Y^i_{,j} \, d\textup{vol}_{\og} + \int_{A_{\rho} } \sum_{i,j,k, p} \og^{jk} \op_{ki} Y^p \overline{\Gamma}_{jp}^i \, d\textup{vol}_{\og}. \label{eq:angularmomentum}
\end{align}
In the first equality, we use the constraint equation $\mbox{div}_{\og}  \op = 0$. Then because $Y$ is Killing (with respect to the Euclidean metric), 
\begin{align*}
		 \sum_{i,j,k} \og^{jk} \op_{ki} Y^i_{,j} =\sum_{i,j,k} (\og^{jk}-\delta^{jk}) \op_{ki} Y^i_{,j}.
\end{align*}
By \eqref{eq:gasymptotics}, \eqref{eq:piasymptotics}, the above identity, and that $Y$ is Killing, we have
\begin{align} \label{eq:firstterm}
\begin{split}
		 \sum_{i,j,k} \og^{jk} \op_{ki} Y^i_{,j} &= -\sum_{i,j,k} \sigma_{jk} \tau_{ki}Y^i_{,j} \\
		 &\quad + \frac{1}{r^3} \sum_{i,j,n}\sigma_{jn}(B_nx_i + B_i x_n)  Y^i_{,j}+ O(r^{-3-q}).
\end{split}
\end{align}
Then by \eqref{eq:exsigma} and \eqref{eq:extau}, the first line vanishes:
\begin{align*}
		\sum_{i,j,k}\sigma_{jk} \tau_{ki}Y^i_{,j} &= \frac{\alpha \beta}{r^3}\sum_{i,j,k}\left( \frac{x_j x_k}{r^2} - \frac{1}{2} \delta_{jk} \right) \frac{x_ix_k}{r^2} Y^i_{,j} = \frac{\alpha \beta}{2 r^4}\sum_{i,j}  \frac{x_i x_j }{r}Y^i_{,j}\\
		& = \frac{\alpha \beta}{2 r^4}\sum_{i}  x_i \frac{\partial Y^i}{\partial r} =  \frac{\alpha \beta}{2 r^4}\frac{\partial}{\partial r} \left(\sum_{i} x_iY^i \right)=0,
\end{align*} 
where we use that $Y$ is a rotation vector field and hence $\sum_i x_i Y^i =0$. The second line in \eqref{eq:firstterm} is 
\[
		-\frac{\alpha}{r^4} \sum_{i,j} Y^i_{,j} B_j x_i  + O(r^{-3-q}).
\]
 
Because $\og = \delta + O(r^{-1})$, the second integral in \eqref{eq:angularmomentum} is
\begin{align*}
		&\int_{A_{\rho} } \sum_{i,j,p} \og^{jk} \op_{ki} Y^p \overline{\Gamma}_{jp}^i \, d\textup{vol}_{\og} = \frac{1}{2} \int_{A_{\rho} } \sum_{i,j, p} \op_{ij} Y^p \og_{ij,p} \, dx + O(1)\\
		&= \frac{1}{2} \int_{A_{\rho} } \sum_{i,j, p}\frac{1}{r^3} \left[ -2 B_i x_j + \sum_n (B_n x_n)  \delta_{ij} \right]Y^p \sigma_{ij,p} \, dx \\
		 &\quad+ \frac{1}{2} \int_{A_{\rho} } \sum_{i,j, p} \tau_{ij} Y^p \sigma_{ij,p} \, dx  + O(1).
\end{align*}
Then the lemma follows by substituting $\sigma$ and $\tau$ by \eqref{eq:exsigma} and \eqref{eq:extau},
\qed
\end{pf}
\begin{cor} \label{co:ang}
If we choose 
\[
	\alpha = \frac{x_1^2}{r^2}, \quad \beta= \frac{x_1x_3}{r^2}, \quad \mbox{ and } \quad Y = x_3 \partial_1 - x_1 \partial_3,  
\]
then $\alpha, \beta \in C^{\infty}(S^2)$ in spherical coordinates, and $\beta$ violates the RT condition.  Moreover, the integral of the angular momentum with respect to $Y$ diverges. 
\end{cor}
\begin{pf}
By Lemma \ref{lemma:angularmomentum} and the straightforward computations, 
\[
	\frac{1}{4} \int_{A_{\rho} } \sum_p\frac{\alpha_{,p}\beta}{r^3}Y^p \, dx =\frac{1}{2} \int_{A_{\rho}} \frac{x_1^2 x_3^2}{ r^7} \, dx \rightarrow \infty \quad \mbox{as } \rho \rightarrow \infty.
\]
The other terms in \eqref{eq:ang} vanish because $\alpha$ is an even function and $Y$ is odd. 
\qed
\end{pf}
\begin{rmk}
The result can be thought as gravity analogue of the addition to a Newtonian system of the mass distributed from a finite radius to infinity whose motion imposes infinite amount of angular momentum about the rotation axis.
\end{rmk}
\end{exang}

\begin{exctm}
We construct an explicit example of an asymptotically flat manifold  whose center of mass diverges. 

\begin{lemma} \label{lemma:scalar}
For any asymptotically flat Riemannian metric $g = \delta + O_2(r^{-1})$, the scalar curvature has the asymptotics:
\begin{align}
	R_g(x) = \sum_{i,j}( g_{ij,ij} - g_{ii,jj}) + E_g + O(r^{-5}),  \quad \mbox{when  $r$ large}, \label{eq:scalar} 
\end{align}
where 
\begin{align*}
	E_g =& - \sum_{i,j,l} (g_{il} - \delta_{il} )(2 g_{ij,lj} - g_{il,jj} - g_{jj,li})\\
	&+\sum_{i,j,l} \left(-g_{jl,j} g_{il,i} + g_{jl,j}g_{ii,l}+ \frac{3}{4} g_{ij,l}^2 - \frac{1}{4} g_{jj,l} g_{ii,l} - \frac{1}{2} g_{ij,l} g_{il,j} \right). 
\end{align*}
\end{lemma}
\begin{pf}
For any Riemannian metric $g$, over a coordinate chart, we have
\begin{align*}
	R_g &= g^{ik} g^{jl} g(\nabla_{\partial_j} \nabla_{\partial_i} \partial_k - \nabla_{\partial_i} \nabla_{\partial_j} \partial_k, \partial_l) \\
	&= g^{ik}\big[ (\Gamma^j_{ik,j} - \Gamma^j_{jk,i}) + (\Gamma^m_{ik} \Gamma^j_{jm} - \Gamma^m_{jk} \Gamma^j_{im}) \big].
\end{align*}
We choose the asymptotically flat coordinates. Using the property that $g_{ij} = \delta_{ij} + O_2(r^{-1})$ and by direct computations, we have the following asymptotics:
\begin{align*}
	&g^{ik} (\Gamma^j_{ik,j} - \Gamma^j_{jk,i}) \\
	&=  \frac{1}{2} g^{ik} \Big\{ \big[g^{jl} (g_{il,k}+ g_{kl,i} - g_{ik,l} ) \big]_{,j} - \big[ g^{jl} ( g_{jl,k} + g_{kl,j} - g_{jk,l} )\big]_{,i}\Big\}\\
	& =\sum_{i,j,l} \big[- g_{jl,j} g_{il,i} + \frac{1}{2} g_{jl,j} g_{ii,l} + \frac{1}{2} g_{jl,i}^2\big]\\
	&\quad +\sum_{i,j} (g_{ij,ij} - g_{ii,jj}) - (g_{ik} -\delta_{ik}) (2 g_{ij,kj} - g_{ik,jj} - g_{jj,ki}) + O(r^{-5}),
\end{align*}
where in the second equality, we use $g^{jl}_{,j} = - g_{jl,j} + O(r^{-3})$. By the straight forward computations and the definition of the Christoffel symbols, 
\begin{align*}
	 &g^{ik} \big[(\Gamma^m_{ik} \Gamma^j_{jm} - \Gamma^m_{jk} \Gamma^j_{im}) \big] \\
	 &= \sum_{i,j,l} \left( \frac{1}{2}g_{jl,j} g_{ii,l}  - \frac{1}{4} g_{jj,l} g_{ii,l} - \frac{1}{2}  g_{ij,l}g_{il,j} + \frac{1}{4} g_{ij,l}^2\right)+ O(r^{-5}).
\end{align*}
Combining above identities, we derive \eqref{eq:scalar}.
\qed
\end{pf}

Let $(\og, \op)$ be an asymptotically flat manifold of the asymptotics \eqref{eq:gasymptotics} and \eqref{eq:piasymptotics}. If we let $(g,\pi) = (\delta,0)$ and $\tau =0$, from the proof of Theorem \ref{thm:main}, we have $\op = 0$ and hence $\og$ satisfies the constraint equation $R_{\overline{g}} = 0$.
\begin{prop}
Let $\og$ satisfy \eqref{eq:og} with $R_{\og}=0$. Its center of mass is equal to, for $l=1,2,3$,
\begin{align} \label{eq:ctm}
	\C_l = \frac{1}{16 \pi E} \lim_{\rho \rightarrow \infty} \int_{A_{\rho}} - x_l \left[\frac{3}{2} \frac{A \alpha}{r^4} + \frac{ \alpha}{2r^2} \alpha_{,jj} - \frac{33}{8} \frac{ \alpha^2} {r^4} + \frac{3}{8}\frac{ | \nabla \alpha |^2 }{ r^2} \right]\, dx + O(1).
\end{align}
\end{prop}
\begin{pf}
Let $A_{\rho}= \{ x \in \mathbb{R}^3 : \rho_0 \le |x| \le \rho \}$ for some fixed $\rho_0$. Then by the divergence theorem
\begin{align*}
	 &\int_{\partial A_{\rho}} \left[ x_l \sum_{i,j}(\og_{ij,i} - \og_{ii,j} ) \frac{x_j}{r}-  \sum_i \left(\og_{il} \frac{x_i}{r} - \og_{ii} \frac{x_l}{r} \right) \right]\, d S\\
	 & = \int_{A_{\rho}} x_l \sum_{i,j} (\og_{ij,ij} - \og_{ii,jj}) \, dx.
\end{align*}
Using the identity \eqref{eq:scalar},  
\[
	x^l \sum_{i,j}(\og_{ij,ij} - \og_{ii,jj}) = x^l R_{\og} - x^l E_{\og} + O(r^{-4}).
\]
The first term of the scalar curvature vanishes, and the third term is integrable over $A_{\rho}$. It remains to compute  $\int_{A_{\rho}} x^l E_{\og} \, dx$. By Lemma \ref{lemma:scalar}, \eqref{eq:gasymptotics} and direct computations, up to terms of order $O(r^{-4-q})$,
\begin{align*}
		E_{\bar{g}}&=- \frac{A}{r^3}\sum_i \sigma_{ii} + \sum_{i,l} \frac{3A x^i x^l }{r^5} \sigma_{il} -\sum_{i,l,j}\sigma_{il} \left[2 \sigma_{ij,lj} - \sigma_{il,jj} - \sigma_{jj,li} \right]\\
		&\quad + \frac{3}{2} \frac{A^2} {r^4} -\sum_{i,j} \frac{ A x^i}{r^3} \sigma_{jj,i}\\
	&\quad +\sum_{i,j,n}\left[ - \sigma_{jn,j} \sigma_{in,i}+ \sigma_{jn,j} \sigma_{ii,n} +\frac{3}{4} \sigma_{jn,i}^2-\frac{1}{4} \sigma_{jj,n} \sigma_{ii,n} -\frac{1}{2}\sigma_{jn,i} \sigma_{ij,n}\right].
\end{align*}
We then substitute $\sigma$ by \eqref{eq:exsigma}. By the straightforward computations, 
	\begin{align*}
		E_g &= \frac{3}{2} \frac{A^2}{r^4} + \frac{3}{2} \frac{A \alpha}{r^4} + \frac{ \alpha}{2r^2} \alpha_{,jj} - \frac{33}{8} \frac{ \alpha^2} {r^4} + \frac{3}{8}\frac{ | \nabla \alpha |^2 }{ r^2} .
	\end{align*} 
Therefor, \eqref{eq:ctm} follows directly. 
\qed
\end{pf}
\begin{cor}\label{co:ctm}
If we choose 
\[
	\alpha = \frac{x_1}{r}, 
\]
then the first component of the center of mass $\C_1$ is infinity. 
\end{cor}
\begin{pf}
It is easy to check that the term $A\alpha /r^4$ in \eqref{eq:ctm} diverges.
\qed
\end{pf}
\begin{rmk}
This result can be thought as the gravity analogue that adding the rest mass to Newtonian system, and the rest mass is unevenly distributed from a finite radius to spatial infinity. It produces an infinite change in the center of mass of the new system. 
\end{rmk}
\end{exctm}

\subsection{Solutions at the decay rate less than $1$} \label{subsec:2}
We consider another family of $\tau$ satisfying  \eqref{eq:tau}. Let $u$ be any $C^2$ function on $\mathbb{R}^3$. Let 
\[
		\tau_{ij} = (| \nabla u|^2 + u \Delta u ) \delta_{ij} - (u_{i} u_{j} + u u_{ij}).
\]
By direct computations, 
\[
	\sum_i \tau_{ij,i} = 0 \quad \mbox{for } j = 1, 2, 3.
\]
We can choose, for example, $u = \log r$. Then 
\begin{align} \label{eq:tau2}
	\tau_{ij} = \frac{1}{r^2} \delta_{ij} + \frac{ x_i x_j}{ r^4} (2 \log r -1).
\end{align}
Notice that $\tau_{ij} \neq O_1(r^{-2})$  because the logarithmic term. More generally, if we let $u = r^{(1-q)/2}$ for $q<1$, $\tau = O_1(r^{-1-q})$. 

Choosing this particular $\tau$ from \eqref{eq:tau2}, we have another example of a vacuum asymptotically flat manifold whose angular momentum is not defined.
\begin{cor} \label{co:ang2}
Let $(\og, \op)$ satisfy \eqref{eq:gasymptotics} and \eqref{eq:piasymptotics} where $\sigma$ and $\tau$ have the expression \eqref{eq:exsigma} and \eqref{eq:tau2}. Then if 
\[
		\alpha = \tan^{-1} \left( \frac{x_2}{x_1} \right), 
\]
the angular momentum with respect to $Y = -x_2 \partial_1 + x_1 \partial_2$ diverges. 
\end{cor}

\end{document}